\documentclass[10pt,twocolumn,pra]{revtex4-1}
\usepackage{graphicx}
\usepackage{amsthm,amsfonts,amstext,amssymb,amsmath}
\usepackage{latexsym}
\usepackage[utf8]{inputenc}
\usepackage[T1]{fontenc}
\usepackage[usenames,dvipsnames]{color}
\usepackage{braket}
\usepackage{bbold}
\usepackage{placeins}
\usepackage{enumerate}
\usepackage{float}
\usepackage{bm}
\usepackage[pdftex,
pdfauthor={Andreas Elben et al.},
pdftitle={Cross-Platform Verification of Intermediate Scale Quantum Devices}
colorlinks=true,
allcolors=black,hidelinks]{hyperref}

\let\counterwithin\relax
\usepackage{chngcntr}

\renewcommand{\vec}[1]{\mathbf{#1}}

\newcommand*{\ketbra}[2]{\ensuremath{\ket{#1}\bra{#2}}}

\newcommand*{\tr}[2][]{\ensuremath{\textrm{Tr}_{#1}\left[ #2 \right]}}
\newcommand{\id}{\ensuremath{{\mathbb{1}}}}
\newcommand{\Fgm}{\ensuremath{\mathcal{F}_{\textrm{GM}}}}
\newcommand{\Fmax}{\ensuremath{\mathcal{F}_{\textrm{max}}}}
\DeclareMathAlphabet{\mathscr}{OT1}{pzc}{m}{it} 

\graphicspath{{Figures/}}

\begin{document}

	\title[]{Cross-Platform Verification of Intermediate Scale Quantum Devices }
	\author{Andreas Elben\textsuperscript{1,2}, Beno\^it Vermersch\textsuperscript{1,2}, Rick van Bijnen\textsuperscript{1,2}, Christian Kokail\textsuperscript{1,2}, Tiff Brydges\textsuperscript{1,2}, Christine Maier\textsuperscript{1,2}, Manoj K.\ Joshi\textsuperscript{1,2}, Rainer Blatt\textsuperscript{1,2}, Christian F. Roos\textsuperscript{1,2}, Peter Zoller\textsuperscript{1,2}}
		\affiliation{\textsuperscript{1}{Center for Quantum Physics, University of Innsbruck, Innsbruck A-6020, Austria,}}
	\affiliation{\textsuperscript{2}{Institute for Quantum Optics and Quantum Information of the Austrian Academy of Sciences,  Innsbruck A-6020, Austria.}}

\begin{abstract}
	We describe a protocol for cross-platform verification of quantum simulators and quantum computers. 
We show how to measure directly the overlap $\tr[]{\rho_1 \rho_2}$ and the purities  $\tr[]{\rho^2_{1,2}}$, and thus a  fidelity  of two possibly mixed, quantum states $\rho_1$ and $\rho_2$ prepared in separate experimental platforms.  We require only local  measurements in randomized product bases, which are communicated classically.
As a proof of principle, we present the measurement of experiment-theory fidelities for  entangled $10$-qubit quantum states in a trapped ion quantum simulator. 
\end{abstract}

\maketitle

There is an ongoing effort to build intermediate scale quantum devices
involving several tens of qubits~\cite{Preskill2018}.
Engineering 
and physical realization of quantum computers and quantum simulators
are beeing pursued with different physical platforms ranging from atomic and
photonic to solid-state systems. 
Recently, verification procedures \cite{Eisert2019}, such as randomized and cyclic benchmarking~\cite{Emerson2005,Emerson2007,Knill2008,Lu2015,Erhard2019}, and direct fidelity estimation \cite{DaSilva2011,Flammia2011,Lanyon2017} have been  developed, which allow one to compare  an implemented, noisy quantum process (or state)  with a known, theoretical target. 
A key challenge  is the \emph{direct
comparison of a priori unknown quantum states} generated on two devices at different locations and
times by running a specific quantum
computation or quantum simulation, i.e.\  the cross-platform
verification of these experimental quantum devices by means of a fidelity measurement. This will become particularly relevant when
we approach  regimes where eventually a  comparison with classical simulations becomes computationally hard  and thus a direct comparison of quantum machines is needed.  

Our aim is the development of protocols for cross-platform verification by measuring the overlap
of quantum states produced with two different experimental setups, potentially realized on very different physical platforms,
without any prior assumptions on the quantum states themselves. For two pure quantum states, the relevant fidelity is defined as
the overlap $\mathcal{F}_{\textrm{pure}}(\ket{\psi_{1}},\ket{\psi_{2}})=|\braket{\psi_{1}|\psi_{2}}|^{2}$,
where $\ket{\psi_{1}}$ and $\ket{\psi_{2}}$ denote pure states
in Hilbert space $\mathcal{H}$ on devices 1 and 2, respectively. For
mixed states we consider the fidelity \cite{Liang2019}
\begin{align}
\mathcal{F}_{\textrm{max}}(\rho_{1},\rho_{2})=\frac{\tr{\rho_{1}\rho_{2}}}{\max\{\tr{\rho_{1}^{2}},\tr{\rho_{2}^{2}}\}},
\label{eq:eq1}
\end{align}
which measures the overlap between density matrices $\rho_{1}$ and
$\rho_{2}$, respectively, normalized by their purities. Here $\rho_{1}$ ($\rho_{2}$) can refer to the total system, or a subsystem 
of device 1 (2). 
 $\Fmax$ fulfills
the axioms for mixed state fidelities imposed by Josza \cite{Jozsa1994}. 
It can thus be used to verify that, and to which degree, two quantum devices have prepared the same quantum state. We note that the performance of quantum devices has been previously investigated by comparing outcome distributions of a selection of observables   \cite{Linke2017,Greganti2019}. In contrast, we are interested here in specifically measuring the fidelity (1) of  the entire density matrices $\rho_{1}$ and
$\rho_{2}$.
\begin{figure}[t]
	\centering
	\includegraphics[width=0.95\linewidth]{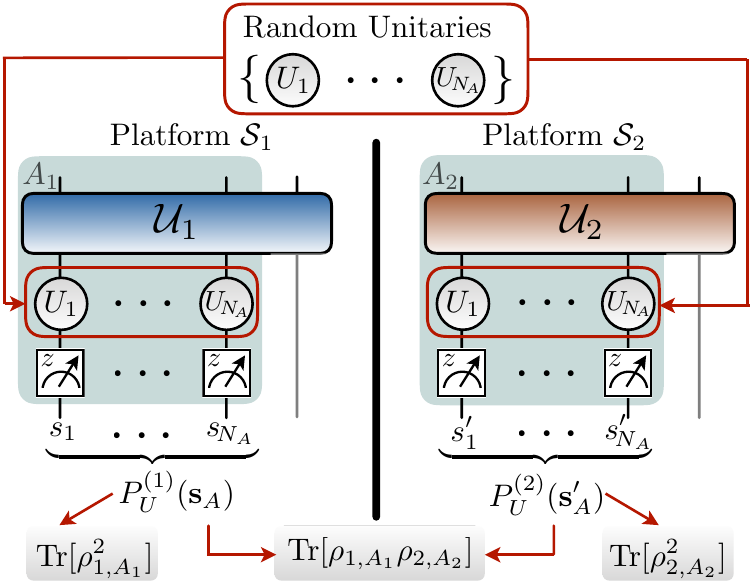}
\caption{\textit{Fidelity estimation with randomized measurements.} We present a protocol to measure the fidelity $\Fmax(\rho_{1,A_1},\rho_{2,A_2})$  of two quantum states described by (reduced) density matrices $\rho_{i,A_i}=\tr[\mathcal{S}_i\setminus A_i]{\rho_i}$ ($i=1,2$): On two platforms $\mathcal{S}_1$ and $\mathcal{S}_2$, the quantum states  $\rho_1$ and $\rho_2$ are prepared with  quantum operations $\mathcal{U}_1$ and $\mathcal{U}_2$, respectively. Randomized measurements are performed on both platforms in (sub)systems $A_1\subseteq \mathcal{S}_1$ and $A_2\subseteq \mathcal{S}_2$ of size $N_A$, implemented with the \emph{same} local random unitaries $U_1\otimes\dots\otimes U_{N_A}$ which are shared via classical communication (red arrows).
From statistical cross- \mbox{(auto-)} correlations  of outcome probabilities $P_U^{(i)}(\mathbf{s}_A)$ ($i=1,2$) the overlap $\tr{\rho_{1,A_1}\rho_{2,A_2}}$ (the purities $\tr{\rho^2_{i,A_i}}$) and thus  $\Fmax(\rho_{1,A_1},\rho_{2,A_2})$   are inferred (see text).}
	\label{fig:Figure1}
\end{figure}
\nopagebreak

The protocol discussed below infers the cross-platform fidelity $\Fmax$ from statistical correlations between randomized measurements performed on the first and second device (see Fig.~\ref{fig:Figure1}).
 While in previous work we obtained R\'{e}nyi (entanglement) entropies, or purities, of reduced density matrices $\tr{\rho_{1,2}^{2}}$, for {\em single systems} from randomized measurements \cite{vanEnk2012,Elben2018,Elben2019} [see the denominator of Eq.~(1)],  we are here interested  in measuring the overlap between density operators of device 1 and 2 from such protocols [see numerator of Eq.~(1)]. In principle, $\Fmax$ can be determined from full quantum state tomography (QST) of systems 1 and 2 \cite{Haffner2005,Gross2010,Riofr2017,Lanyon2017, Torlai2018,Keith2018}.
However, due to the  exponential scaling with the (sub)system size \cite{Gross2010}, this approach is limited to only a few degrees of freedom  \cite{Haffner2005}. Alternative efficient tomographic methods require a specific structure, or {\em a priori} knowledge of the system of interest \cite{Lanyon2017,Torlai2018,Keith2018,Carleo2019}. In contrast, as demonstrated below, the present protocol scales, although exponentially,  much more favorably with the (sub)system size, allowing practical cross-platform verification for (sub)systems involving tens of qubits on state-of-the-art quantum devices \footnote{Our protocol assumes only classical communication between platforms 1 and 2. Existence of a quantum link would allow quantum state transfer and provide, in principle, an efficient quantum protocol for $\tr{ \rho_1 \rho_2}$ \cite{Daley2012,Islam2015,Linke2018}} \nocite{Daley2012,Islam2015,Linke2018}.

In the following, we  first describe the protocol, followed by an analysis of statistical errors and the required number of experimental runs.
Using the data taken in context of Ref.~\cite{Brydges2019}, we demonstrate, as a proof-of-principle, the measurement of experiment-theory fidelities  of  quantum states of $10$ qubits prepared via quench dynamics on a trapped ion quantum simulator. Finally, we present experiment-experiment fidelities of quantum states prepared sequentially on the same experimental platform.

\textit{Protocol. --} As illustrated in Fig.~\ref{fig:Figure1}, we consider two quantum devices consisting of
$N_{1}$ and $N_{2}$ spins ($d$-level systems) realized on  different physical
platforms $\mathcal{S}_{1}$ and $\mathcal{S}_{2}$, and prepared with  quantum operations $\mathcal{U}_1$ and $\mathcal{U}_2$ in two quantum states described by the density matrices $\rho_{1}$ and $\rho_{2}$,
respectively. We denote the reduced density matrices as $\rho_{i,A_{i}}={\rm Tr}_{\mathcal{S}_i\setminus A_{i}}(\rho_{i})$
for (sub) systems $A_{i}\subseteq\mathcal{S}_{i}$ $(i=1,2)$ of identical size $N_{A_{1}}=N_{A_{2}}\equiv N_{A}$. The associated
Hilbert space dimension is $D_{A}=d^{N_{A}}$.

We apply first to both $\rho_{1,A_{1}}$ and $\rho_{2,A_{2}}$ the
\emph{same} random unitary $U_{A}=\bigotimes_{k=1}^{N_{A}}U_{k}$,
defined as a product of local random unitaries $U_{k}$ acting on
spins $k=1,\dots,N_{A}$ (see Fig.~\ref{fig:Figure1}). Here, the $U_{k}$ are sampled independently from
a unitary $2$-design \cite{Gross2007,Dankert2009} defined on the local Hilbert space $\mathbb{C}^d$ and sent via classical communication to both devices (red arrows in Fig.~\ref{fig:Figure1}). We now perform for the first
and second system projective measurements in a standard (computational)
basis $\ket{\vec{s}_{A}}\equiv\ket{s_{1},\dots,s_{N_{A}}}$. Here,
$\vec{s}_{A}$ denotes a string of possible measurement outcomes for
spins $k=1,\dots,N_{A}$. Repeating these measurements  
for fixed $U_{A}$
provides us with estimates of the probabilities $P_{U}^{(i)}(\vec{s}_{A})={\rm Tr}_{A_i} [{U_{A}\,\rho_{i,A_i}\,U_{A}^{\dagger}\ketbra{\vec{s}_{A}}{\vec{s}_{A}}}]$
 for $i=1,2$ (see Fig.~\ref{fig:Figure1}). In a second step, this procedure is repeated for many different random unitaries $U_{A}$.

Finally,  we estimate 
 the density matrix 
overlap $\tr[]{\rho_{1,A_{1}}\rho_{2,A_{2}}}$ from  second-order \emph{cross-correlations} between the
two platforms via
\begin{align}
\tr[]{\rho_{i,A_{i}}\rho_{j,A_{j}}}\!=\! \label{eq:ovl} 
d^{N_{A}}\!\sum_{\mathbf{s}_{A},\mathbf{s}_{A}'}\!(-d)^{-\mathcal{D}[\mathbf{s}_{A},\mathbf{s}_{A}']}\;\overline{P_{U}^{(i)}(\mathbf{s}_{A})P_{U}^{(j)}(\mathbf{s}_{A}')}.
\end{align}
with $i\!=\!1$, $j\!=\!2$.
This is proven in the Supplemental Material \cite{SM}
,  Appendix \ref{app:Proof},using the properties of unitary $2$-designs, thus generalizing~\cite{Elben2019} to cross-platform settings.
  Here, $\overline{\vphantom{a}\dots}$ denotes the ensemble average over random unitaries of the form $U_A$. The Hamming distance $\mathcal{D}[\mathbf{s}_{A},\mathbf{s}_{A}']$
between two strings $\mathbf{s}_{A}$ and $\mathbf{s}_{A}'$
 is defined
as the number of spins  where $s_{k}\neq{s}'_{k}$,
i.e.\ $\mathcal{D}[\mathbf{s}_{A},{\mathbf{s}_{A}}']\equiv|\left\{ k\in \{1,\dots, N_A\}\,|\,s_{k}\neq{s}'_{k}\right\} |$. 
 The purities $\tr[]{\rho_{1,A_1}^{2}}$ and $\tr[]{\rho_{2,A_2}^{2}}$ for the first and second subsystem are obtained by setting in Eq.~(\ref{eq:ovl}) $i\!=\!j\!=\!1$ and $\!i\!=\!j\!=\!2$, respectively, i.e.~as second-order
\emph{auto-correlations} of the probabilities  $P_{U}^{(i)}(\mathbf{s}_{A})$
and $P_{U}^{(i)}(\mathbf{s}_{A}')$ ~\cite{Brydges2019,Elben2019}. 

We emphasize that the above protocol to measure the cross-platform fidelity of  two quantum states requires only classical communication of  random unitaries and measurement outcomes between the two platforms, with the experiments  possibly taking place at very different points in time and space. In its present form, the protocol requires, or assumes no prior knowledge of the quantum states. These states can be mixed states, and refer to subsystems, allowing in particular a comparison of subsystem fidelities for various sizes.  We note that our protocol can  be used to perform fidelity estimation towards known target theoretical states, as an experiment-theory comparison (see below). In this setting, and when the  `theory state’ is pure, direct fidelity estimation protocols have been developed~\cite{Flammia2011,DaSilva2011}, that can be more  efficient for certain well-conditioned states, which are supported on a small number of multiqubit Pauli operators.

\textit{Scaling of the required number of experimental runs --}
In practice, a statistical error  of the estimated fidelity arises from a finite number of projective measurements $N_M$ performed per random unitary and a finite number $N_U$ of random unitaries used  to infer overlap and purities via Eq.~(\ref{eq:ovl}). 
Experimentally  relevant is, therefore,  the  scaling of the total number of experimental runs $N_M N_U$ (the measurement budget), which are required to reduce this statistical error below a fixed value $\epsilon$, for $N_A$ qubits. In addition, there is  the optimal allocation of resources,  $N_U$ and $N_M$, for a given measurement budget $N_M N_U$.

\begin{figure}
	\centering
	\includegraphics[width=1.\linewidth]{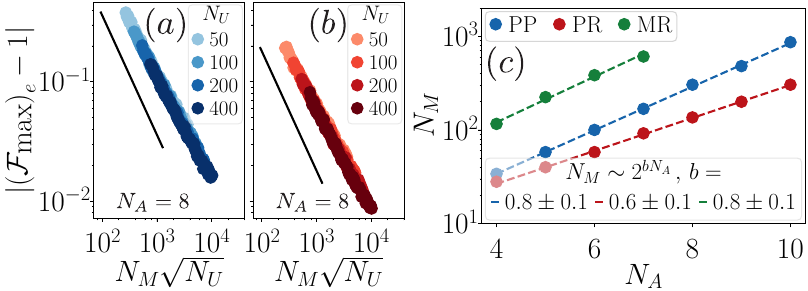}
	\caption{\textit{Scaling of the required number of measurements.} (a,b) Average statistical error $|(\Fmax(\rho_A,\rho_A))_e -1 |$  as a function of the number of measurements $N_M$ per random unitary for various $N_U$ (darkness of colors). The state $\rho_A$ of $N_A=8$ qubits ($d=2$)  is taken to be  (a) a pure product state (PP)  and (b) a pure Haar random state (PR).  Black lines are guides for the eye $\sim 1/(N_M \sqrt{N_U})$. (c) Scaling of the minimal number of required measurements $N_M$  to estimate $(\Fmax(\rho_A,\rho_A))_e$ up to a fixed   statistical error of $0.05$ as a function of the number of qubits $N_A$, for fixed $N_U=100$. The mixed random states (MR) are obtained from tracing out $3$ qubits from Haar random states of $N_A+3$ qubits.}
	\label{fig:StatErr}
\end{figure}

In Fig.~\ref{fig:StatErr} we present numerical results   for the average statistical error as  a function of $N_M$ and $N_U$, and infer the scaling of the measurement budget  with \mbox{(sub)system} size $N_A$. For simplicity, we  assume that the  target fidelity $\Fmax(\rho_{1,A_1},\rho_{2,A_2})$ for the two states $\rho_{1,A_1}$ and $\rho_{2,A_2}$ is known and  analyze the scaling of the statistical error $|\left[ \Fmax(\rho_{1,A_1},\rho_{2,A_2}) \right]_e - \Fmax(\rho_{1,A_1},\rho_{2,A_2})| $ of an estimated fidelity $\left[ \Fmax(\rho_{1,A_1},\rho_{2,A_2}) \right]_e$. Focusing on experimentally relevant system sizes, we simulate experiments by applying  $N_U$ random unitaries to  $\rho_{1,A_1}$ and $\rho_{2,A_2}$ and sample  independently $N_M$ projective measurements from each state. We then infer an estimation $[\Fmax(\rho_{1,A_1},\rho_{2,A_2})]_e$ of the  fidelity $\Fmax(\rho_{1,A_1},\rho_{2,A_2})$ using Eq.~(\ref{eq:ovl}), and calculate - from many  of these numerical experiments - the average statistical error  $|[\Fmax(\rho_{1,A_1},\rho_{2,A_2})]_e -\Fmax(\rho_{1,A_1},\rho_{2,A_2}) |$. In Fig.~\ref{fig:StatErr} we concentrate on the case where the quantum states $\rho_{1,A_1}=\rho_{2,A_2}=\rho_A$ on the two platforms are identical, i.e. the exact fidelity equals $\Fmax(\rho_A,\rho_A)=1$ (for the general case see Appendix \ref{app:StatErros} \cite{SM}).

In Fig.~\ref{fig:StatErr}(a,b),  the average statistical error $|[\Fmax(\rho_A,\rho_A)]_e -1 |$ is shown  as a function of  $N_M$ for  a system of $N_A=8$ qubits ($d=2$) and various  $N_U$ and for two very different types of states $\rho_A$: (a)  pure product states (PP) and (b)  pure (entangled)  Haar random states (PR)  which are obtained by applying a Haar random unitary to a pure product state \cite{SM}.
Our numerical analysis shows that, in the regime $N_M\lesssim D_A $ and $N_U \gg 1$,  $|[\Fmax(\rho_A,\rho_A)]_e -1 | \sim 1/(N_M \sqrt{N_U})$. 
For unit target fidelity, the optimal allocation of the total measurement budget $N_UN_M$ is thus to keep $N_U$ small and fixed \footnote{To allow for estimation of the statistical uncertainty of estimated fidelity a minimal number of $N_U\gg 1$ is required.}.   

Fixing $N_U=100$, we display  in Fig.~\ref{fig:StatErr}(c) the scaling of the number of projective measurements $N_M$ per unitary required
to determine the fidelity $\Fmax(\rho_A,\rho_A)$  up to an average statistical error $|[\Fmax(\rho_A,\rho_A)]_e -1 |\leq \epsilon$ below  $\epsilon=0.05$.    We find  a scaling  $N_M\sim 2^{bN_A}$  with $b=0.8 \pm 0.1$ for PP and $b = 0.6 \pm 0.1$ for PR states, which persists for tested $\epsilon=0.02,\dots, 0.2$. The  fidelity estimation of PR (entangled) states is thus less prone to statistical errors which we attribute to the fact that fluctuations across random unitaries are reduced due to the mixedness of the subsystems.   A similar scaling, with larger prefactor, is found for a mixed random state (MR), obtained from tracing out $3$ qubits of a random state of $N_A+3$  qubits.  This is directly related to the smaller overall magnitude of numerator and denominator of the fidelity for mixed states [see Eq.~\eqref{eq:eq1}].

We note that the optimal allocation of $N_U$ vs. $N_M$ for given $N_U N_M$ depends on the quantum states, in particular their fidelity and the allowed statistical error $\epsilon$, and is thus a priori not known. In practice, an iterative procedure can be applied in which the allocation of measurement resources $N_U$ vs.\ $N_M$  is stepwise inferred from newly acquired data.  To this end, the expected reductions of the standard error of the estimated fidelity are calculated, upon increasing either $N_U$ or $N_M$, using resampling techniques (see Appendix \ref{app:StatErros} \cite{SM}). Accordingly, $N_U$ and $N_M$ are updated iteratively to maximize the expected decrease of statistical uncertainty, until a predefined value of the estimated error is reached.

In summary, we find  that the presented protocol requires  a total number of experimental runs $N_UN_M\sim 2^{bN_A}$ with $b \lesssim 1$ which is, despite being exponential, significantly less  than full QST with exponents $b\geq 2$ \cite{Gross2010}. For instance, QST via compressed sensing \cite{Gross2010,Riofr2017} would require at least $\mathcal{O}(2^{2N_A})\sim 10^6$ experimental runs for a pure $10$-qubit  state, whereas for our protocol  $10^4$ (PR) to  $10^5$ (PP) experimental runs would be sufficient to obtain a fidelity estimation up to a statistical uncertainty of $0.05$.

\textit{Fidelity estimation with trapped ions --}
In the following,  we present, as proof-of-principle, the measurement of experiment-theory fidelities  and experiment-experiment fidelities 
of  highly-entangled  quantum states prepared via quench dynamics in a trapped ion quantum simulator.  To this end, we use data presented in Ref.~\cite{Brydges2019}. Here, the  entanglement generation after  a quantum quench with  the $XY$-Hamiltonian
\begin{equation}
H_{\mathrm{XY}} = \hbar\sum_{i<j}J_{ij}(\sigma^{+}_{i}\sigma^{-}_{j}+\sigma^{-}_{i}\sigma^{+}_{j}) + \hbar B \sum_{i}\sigma^{z}_{i}
\label{eq:XY-Hamiltonian}
\end{equation}
was experimentally monitored, with  $\sigma_i^{z}$  the third spin-$1/2$ Pauli operator, $\sigma_i^{+}(\sigma_i^{-})$ the spin-raising (lowering) operators acting on spin $i$, and $J_{ij} \approx J_0/ \lvert{i-j}\rvert^{\alpha}$ the coupling matrix with an approximate power-law decay $\alpha\approx 1.24$ and $J_0=420 s^{-1}$. The initial N\'{e}el-state, $\rho_{E}(0)\approx|\psi\rangle\langle\psi|$ with $|\psi\rangle=|0,1,0,\dots,1\rangle$ for $N=10$ ions, was time-evolved under $H_{\mathrm{XY}}$ into the state $\rho_E(t)$.  
Subsequently, randomized measurements were performed and, from statistical auto-correlations of the outcome probabilities $P_U^{(E)}(\mathbf{s}_A)$, purity and second-order R\'{e}nyi entropy of $\rho_E(t)$ (and of  density matrices of arbitrary subsystems), were inferred.  In total, $N_U=500$ random unitaries were used  and $N_M=150$ projective measurements per random unitary were performed.
For further experimental details, see Ref.~\cite{Brydges2019}.

To numerically simulate the experiment and obtain a corresponding theory state $\rho_T(t)$, we perform  exact diagonalization   to simulate   unitary dynamics or exactly  solve a master equation to include decoherence effects. Subsequently, the  $N_U=500$ random unitaries which have been employed in the experiment are applied to $\rho_T(t)$ and the occupation probabilities $P_U^{(T)}(\mathbf{s})$  are calculated exactly for each random unitary.

\begin{figure}[t]
	\centering
	\includegraphics[width=0.99\linewidth]{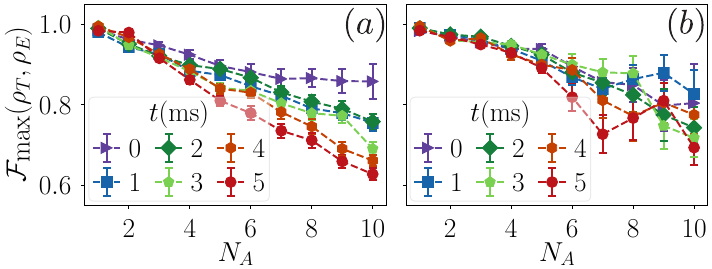}
	\caption{\textit{Experiment-Theory verification in a trapped ion quantum simulator.} Measured fidelities $\Fmax(\rho_E,\rho_T)$ as  a function of partition size $N_A$ (total system $10$ qubits) for states $\rho_E$ evolved with $H_{XY}$ ($J_0= 420 s^{-1},\alpha= 1.24$) for various times; experimental data from \cite{Brydges2019}. Theory states $\rho_T$ are obtained with (a)   unitary dynamics and  including   (b) decoherence effects  (see text).  In both panels, $N_U=500$ and $N_M=150$. Error bars are obtained with Bootstrap resampling \cite{Efron1983}. Dashed lines are guides for the eye.}
	\label{fig:TheoExp}
\end{figure}

In Fig.~\ref{fig:TheoExp}(a,b), experiment-theory fidelities  $\Fmax(\rho_{E,A},\rho_{T,A})$ of reduced states of connected partitions $[1\rightarrow N_A]$ are displayed as  a function  $N_A$ for various times after the quantum quench. For (a) theory states are calculated by simulating unitary dynamics and for (b) we additionally include decoherence effects, inherent to the state preparation (imperfect initial state preparation, spin-flips and dephasing noise) and  the measurement process (depolarizing noise during the random measurement) \cite{Brydges2019}. 
In both cases, we find a single qubit fidelity being constant in time and close to unity. With increasing subsystem size and time, the estimated fidelities tend to decrease.
Remarkably, we  find theory-experiment fidelities  (a) $\gtrsim 0.6$  [(b) $\gtrsim 0.7$]   even at late times  $T=5\textrm{ms}$, when  the system  has undergone complex many-body dynamics and is highly entangled \cite{Brydges2019}.

We observe in Fig.~(3) a decrease of the estimated  fidelity with system size already at $t=0\,\text{ms}$; despite that the initial N\'eel state can be prepared and, being a simple product state, directly verified (preparation fidelity $\gtrsim 0.97$ for $N_A=10$). Thus,  we attribute the decrease of the estimated theory-experiment fidelity mainly to experimental imperfections in the implementation of the randomized measurements, of two types: (i) unitary errors in the form of random under- or overrotations, i.e.\ a mismatch between  the random unitaries applied in experiment and theory and  (ii) decoherence in the form of local depolarizing noise.  While (i) decreases the estimated density matrix overlap, and thus fidelity, in both cases presented in Fig.~\ref{fig:TheoExp}, (ii) is taken into account into the theory state for panel (b) and thus the estimated fidelities are larger than in (a). We emphasize that both sources of imperfections decrease the estimated fidelity and do not lead to false positives and refer for a detailed error modeling and further experimental investigations to Appendix \ref{app:Imperfections} \cite{SM}.

\begin{figure}[t]
	\centering
	\includegraphics[width=0.99\linewidth]{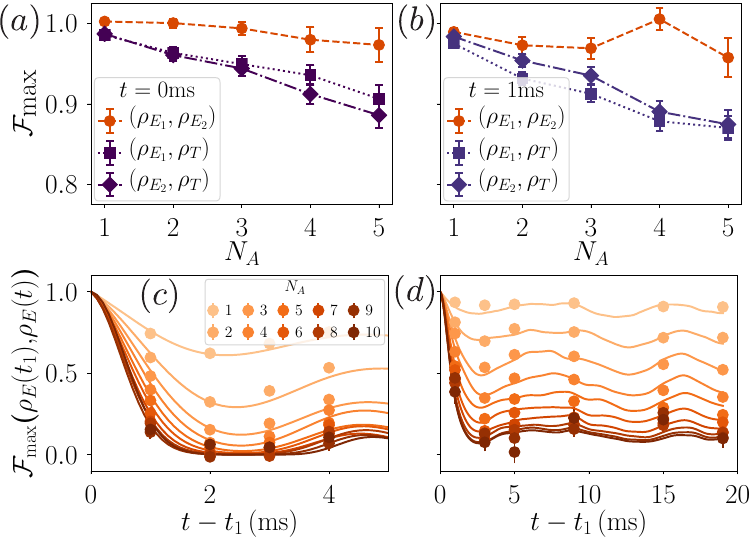}
	\caption{\textit{Experiment self-verification  in a trapped ion quantum simulator.} (a,b) Estimated fidelities $\Fmax$  of two reduced states $\rho_{E_1}$ and $\rho_{E_2}$  prepared sequentially in the same experiment as  a function of partition size, $[1\rightarrow N_A]$. The states $\rho_{E_1}$ and $\rho_{E_2}$  are (a) two N\'{e}el states which have been (b) time-evolved   under $H_{XY}$ ($J_0= 420 s^{-1},\alpha= 1.24$) to $t=1\,\textrm{ms}$; experimental data from \cite{Brydges2019}.	Experiment-theory fidelities are obtained by simulating unitary dynamics (see text). Dashed lines are guides for the eye.   (c,d) Measured fidelities $\Fmax\bm{(}\rho_E(t_1),\rho_E(t)\bm{)}$ for states time evolved with $H_{XY}$ as  a function of the time difference $t-t_1$  ($t_1=1\,\textrm{ms}$)  for (c) a  clean system  and (d)  with additional disorder (see text). Different colors refer to different partitions $[1\rightarrow N_A]$. Lines show theory simulations including decoherence effects (see text). In all panels, error bars are estimated with Bootstrap resampling  \cite{Efron1983}.}
	\label{fig:ExpExp}
\end{figure}

As a first step towards the cross-platform verification of two quantum devices, we now present   experiment-experiment fidelities of  quantum states prepared sequentially in the same experiment. To this end, we divide  the data obtained in Ref.~\cite{Brydges2019} into two parts, from now on called experiment $E_1$ and experiment $E_2$, each consisting of measurement outcomes for the same $N_U=500$  random unitaries and $N_M=75$ measurements per random unitary. Using Eq.~\eqref{eq:ovl}, we calculate  overlap and  purities, and from this the fidelity $\Fmax(\rho_{E_1}(t),\rho_{E_2}(t))$. 
In Fig.~\ref{fig:ExpExp}(a,b), the experiment-experiment and theory-experiment fidelities are displayed as  a function of subsystem size for $t=0,1\,\textrm{ms}$. 
In comparison to theory-experiment fidelities, experiment-experiment fidelities are higher for both $t=0\,\textrm{ms}$ and $t=1\,\textrm{ms}$.
We conclude that the random unitaries are reproducibly prepared in the experiment, with a systematic mismatch (unitary error) compared to the ones on the classical computer.

Finally, we illustrate our method in Fig.~\ref{fig:ExpExp}(c,d) by the measurement of $\Fmax$ of two quantum states evolved for different times. We consider in Fig.~\ref{fig:ExpExp}(c) the clean system, governed by $H_{XY}$, and Fig.~\ref{fig:ExpExp}(d) the case where additional on-site disorder $H_{\text{tot}}=H_{XY}+\sum_j \delta_j \sigma^z_j$, with $ \delta_j   $ sampled uniformly from  $[-3J_0,3J_0]$, is added.
We find that for the clean system the fidelity decays quickly as a function of the subsystem size and time difference, resembling the complex, ergodic dynamics in the interacting many-body system. On the contrary, for the disordered system the fidelity stays, after an initial short-time decay, approximately constant, and at a finite value even for large (sub)systems. Our results are thus consistent with localization phenomena, characterized through the system's memory of earlier time and slow dynamics, as also studied with out-of-time order correlators \cite{Serbyn2017,Fan2016,Chen2017,Huang2017}, also  accessible with randomized measurements \cite{Vermersch2019}.

\textit{Conclusion --} We have presented a protocol to perform cross-platform verification of quantum devices by direct fidelity measurements, requiring only classical communication and significantly fewer measurements than full quantum state tomography. Extrapolating the numerically extracted scaling laws for the required number of experimental runs, we  expect it to be applicable in state-of-the-art quantum simulators and computers with high repetition rates for (sub)systems consisting of a few tens of qubits. In larger quantum systems, it gives access to the fidelities of all possible subsystems up to a given size --  determined by the accepted statistical error and the measurement budget -- and thus enables a fine-grained comparison of large quantum systems. Furthermore, we expect that adaptive sampling  techniques have the potential to reduce the measurement cost, in particular when knowledge over the quantum states of interest is taken into account.

\begin{acknowledgments}
We thank B. Kraus, A. Browaeys, J. Bollinger, Z. Cian, J. Emerson, S. Glancy, M. Hafezi, R. Kaubrügger, A. Kaufmann, A. Gorshkov, D. Leibfried, C. Monroe and his group, T. Monz, D. Slichter, A. Rey, A. Wilson, and J. Ye for discussions. The project has received funding from the European Research Council (ERC) under the European Union’s Horizon 2020 research and innovation programme (Grant Agreement No. 741541), and from the European Union’s Horizon 2020 research and innovation programme under Grant Agreement No. 817482 (Pasquans) and No. 731473 (QuantERA via QTFLAG). Furthermore, this work was supported by the Simons Collaboration on Ultra-Quantum Matter, which is a grant from the Simons Foundation (651440, P. Z.). We acknowledge support by the ERC Synergy Grant UQUAM and by the Austrian Science Fund through the SFB BeyondC (F71). Numerical simulations were realized with QuTiP.
\end{acknowledgments}

\section*{Author contributions}
A.E., B.V., R.v.B., C.K. and P.Z.\ developed the protocol. T.B., C.M., M.J., R.B., C.F.R.\ provided the experimental data.

\bibliographystyle{apsrev4-1}
%


\clearpage
\appendix
\counterwithin{figure}{section}

{
	\centering \bf \Large
	Supplemental Material\vspace*{0.25cm}\\
	\vspace*{0.0cm}
}

\section{Proof of Equation 2}
\label{app:Proof}

In this section, we prove Eq.\ (2) of the main text (MT), relating the overlap $\tr[]{\rho_{i,A_i} \rho_{j,A_j}}$ of the density matrices $\rho_{i,A_i}$ and  $\rho_{j,A_j}$, defined in the two subsystems $A_i$ and $A_j$ consisting of $N_{A_i}=N_{A_j}=N_A$ qudits, to statistical cross-correlations of randomized measurements. 
Here, randomized measurements on both subsystems are implemented with the same local random unitaries of the form $U_A=\bigotimes_{k=1}^{N_A} U_k$, $U_k$ sampled for $k=1,\dots,N_A$ independently from a unitary $2$-design defined on the local Hilbert space $\mathbb{C}^d$. \textcolor{black}{Here, a unitary $k$-design is an ensemble of unitary random matrices which approximates  Haar random unitaries in the sense that the ensemble average for all polynomials of random unitaries up to order $k$ agrees with the average of the Haar random unitaries (see for example  Ref. \cite{Dankert2009} and also \cite{Elben2019} for formal definitions). Furthermore,  Haar random unitaries are random unitary matrices sampled from  the unique probability measure on the space of unitary matrices which is invariant under arbitrary unitary transformations; this measure is called the Haar measure and the ensemble of Haar random unitaries is also called the circular unitary ensemble  \cite{Watrous2018}.}

To prove Eq.~(2) MT, we first rewrite its right-hand-side $\mathcal{R}$, representing a weighted sum of  cross-correlations of outcome probabilities, as the expectation value of an operator $ {O}$ acting on the joint Hilbert space $\mathcal{H} \otimes \mathcal{H}$ of both subsystems $A_i$ and $A_j$, $\mathcal{H}=(\mathbb{C}^d)^{N_A}$.
Using the linearity of the trace operation and of the average over random unitaries,  we find
\begin{align}
\mathcal{R}&=d^{N_{A}}\!\sum_{\mathbf{s}_{A},\mathbf{s}_{A}'}\!(-d)^{-\mathcal{D}[\mathbf{s}_{A},\mathbf{s}_{A}']}\;\overline{P_{U}^{(i)}(\mathbf{s}_{A})P_{U}^{(j)}(\mathbf{s}_{A}')} \nonumber \\
&= \tr[]{\;\overline{U_A^\dagger \otimes U_A^\dagger  O U_A \otimes U_A} \; \rho_{i,A_i} \otimes \rho_{j,A_j}} \nonumber\\
&= \tr[]{ \; \bigotimes_{k=1}^{N_A} \, \overline{U_k^\dagger \otimes U_k^\dagger  O_k U_k \otimes U_k} \;  \rho_{i,A_i} \otimes \rho_{j,A_j}}.
\label{eq:proof1}
\end{align}
Here,  we defined ${O}\equiv \bigotimes_{k=1}^{N_A} {O}_k$ with
\begin{align}
{O}_{k}\equiv d\!\sum_{{s}_{k},{s}_{k}'}\!(-d)^{-\mathcal{D}[{s}_{k},{s}_{k}']} \ket{{s}_k}\bra{{s}_k}\otimes\ket{{s}'_k}\bra{{s}'_k}
\end{align}
acting on the joint Hilbert space $\mathbb{C}^d \otimes \mathbb{C}^d$ of the $k$-th qudits in $A_i$ and $A_j$.  To arrive at the last line of Eq.~\eqref{eq:proof1}, we used the independence of the random unitaries $U_k$ applied to different qudits $k$.

To evaluate the ensemble average $\overline{U_k^\dagger \otimes U_k^\dagger  O_k U_k \otimes U_k}$, we use the Weingarten calculus of Haar random unitaries \cite{Watrous2018,Roberts2017}. One finds \cite{Elben2019}
\begin{align*}
&\overline{U_k^\dagger \otimes U_k^\dagger  O_k U_k \otimes U_k} \\
& \qquad = \frac{1}{d^2-1} \left(\tr[]{  O_k}  - \frac{1}{d} \tr[]{ \mathbb{S}_k  O_k} \right) \id_2  \nonumber \\ &\qquad \quad +  \frac{1}{d^2-1} \left(\tr[]{ \mathbb{S}_k O}  - \frac{1}{d} \tr[]{  O_k} \right)  \mathbb{S}_k\\
&\qquad  = \mathbb{S}_k
\end{align*}
with the swap operator  $\mathbb{S}_k= \sum_{s_k,s_k'}\ket{s_k} \bra{s_k'} \otimes  \ket{s_k'}\bra{s_k}$. 
Finally, we thus obtain
\begin{align}
\mathcal{R} = \tr[]{ \; \bigotimes_{k=1}^{N_A}  \mathbb{S}_k\;  \rho_{i,A_i} \otimes \rho_{j,A_j}} = \tr[]{ \rho_{i,A_i} \rho_{j,A_j}}
\end{align}
proving Eq.~(2) MT.

\section{Geometric mean fidelity}
\label{app:GMFidelity}

In the MT, we concentrate on the estimation of the fidelity $\Fmax(\rho_1,\rho_2)=\tr[]{\rho_1 \rho_2}/\max\{\tr[]{\rho_1^2},\tr[]{\rho_1^2}\}$ of two quantum states $\rho_1$ and $\rho_2$ which fulfills all axioms imposed by Josza \cite{Jozsa1994}. However, the definition of a mixed state fidelity is not unique, and a variety of different approaches exists \cite{Liang2019}. With the presented protocol based on statistical correlations of randomized measurements, we can infer any mixed state fidelity which is solely a function of overlap   $\tr[]{\rho_1 \rho_2}$  and purities $\tr[]{\rho_1^2}$ and $\tr[]{\rho_2^2}$.
As an example, we consider in this section the fidelity
\begin{align}\label{eq:Fgm}
\Fgm(\rho_1,\rho_2) = \frac{\tr[]{\rho_1 \rho_2}}{\sqrt{\tr[]{\rho_1^2} \tr[]{ \rho_2^2}}}
\end{align}
which is obtained by normalizing the overlap $\tr[]{\rho_1 \rho_2}$ with the geometric mean of the purities $\tr[]{\rho_1^2}$ and $\tr[]{\rho_2^2}$. Clearly, it follows that $ \Fgm(\rho_1,\rho_2) \geq  \Fmax(\rho_1,\rho_2)$, with equality for states with identical purity. We further note that, due to the symmetric normalization, $\Fgm$ is  robust against certain types of decoherence. For instance with $D_A=d^{N_A}$ and $\rho'_1=\lambda \rho_1 + (1-\lambda) \id/D_A$ obtained from $\rho_1$ with global dephasing of arbitrary strength $\lambda$,  we have $\Fgm(\rho_1,\rho_2)=\Fgm(\rho'_1,\rho_2)+\mathcal{O}(D^{-1})$ (see Appendix \ref{app:Imperfections} for more details on decoherence effects). 

\begin{figure}
	\centering
	
	\includegraphics[width=0.99\linewidth]{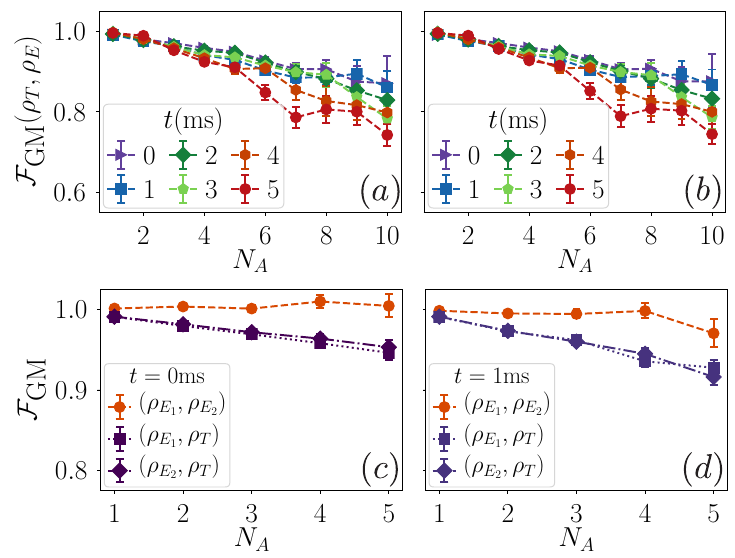}
	\caption{\textit{Theory-Experiment and Experiment self-verification in a 10-qubit system.} (a,b) Estimated theory-experiment fidelity $\Fgm$  of a N\'{e}el state, time-evolved under $H_{XY}$ ($J_0= 420 s^{-1},\alpha= 1.24$), for connected \mbox{(sub)systems} $[1\rightarrow N_A]$ at different times. Experimental data is taken from \cite{Brydges2019}.	Theory states $\rho_T(t)$ have been obtained by simulating unitary dynamics (a) [and also (c,d)] and  taking into account decoherence effects (b). In panels (c,d), additionally experiment-experiment fidelities of two quantum states $\rho_{E_1}$ and $\rho_{E_2}$ obtained sequentially in the same experiment are shown. Here, the experimental data \cite{Brydges2019} has been divided into two parts $E_1$ and $E_2$. In all panels, error bars  for the fidelities are estimated with Bootstrap resampling. $N_U=500$ random unitaries have been used in experiment and theory,  $N_M=150$ (a,b)  [$N_M=75$ (c,d)] measurements per random unitary are performed in the experiment(s).  \textcolor{black}{ Lines are guides for the eye.} } 
	\label{fig:ExpExpGM}
\end{figure}

Complementing  the MT, we show in  Fig.~\ref{fig:ExpExpGM}(a,b) the experiment-theory fidelity $\Fgm(\rho_T,\rho_E)$ between  experimental quantum states $\rho_E$ evolved with $H_{XY}$, and the corresponding theoretical simulation. In (a), the theory states are obtained from simulating unitary dynamics, in (b) we take into account decoherence effects (see main text and Ref.~\cite{Brydges2019}). In contrast to $\Fmax$, the fidelity $\Fgm$ does not change significantly between (a) and (b), as it is robust against local depolarizing noise (see Appendix \ref{app:Imperfections}) which is the dominant source of decoherence in the experiment. Furthermore, we show in Fig.~\ref{fig:ExpExpGM}, experiment-experiment fidelities $\Fgm$ of quantum states obtained sequentially in the same experiment. Consistently with the results for $\Fmax$, we find that the experiment-experiment fidelities are larger than theory-experiment fidelities. 

\section{Statistical errors and resampling}
\label{app:StatErros}

In this appendix, we discuss the statistical errors arising from a finite number $N_U$ of random unitaries to estimate the ensemble average and a finite number of measurements $N_M$ per random unitary. We complement the results presented in the main text where we focused on the case of unit target fidelity. We focus mainly on the numerical investigation of experimentally relevant system sizes. \textcolor{black}{We remark that projective measurements in a basis $\{|\mathbf{s}\rangle \}$ on a quantum state $\rho$ are simulated by sampling the measurement outcomes, using a standard random number generator, from the probability distribution specified by probabilities  $\{P(\mathbf{s})= \langle \mathbf{s}|\rho|\mathbf{s}\rangle \}$.}

The appendix is organized as follows. In Subsection \ref{sec:fid_global}, we discuss first the case where the local random unitaries $U_A=\bigotimes_{k=1}^{N_A} U_k$ acting on individual constituents $i$ are replaced by global random unitaries $U_A$ randomizing the entire subsystem $A$. These yield a smaller overall statistical error (see below) and allow for a semi-analytical determination of statistical errors of the fidelity estimation.  In Subsection \ref{sec:fid_local}, we generalize to local unitaries and investigate numerically  the statistical errors, as function of $N_A$, $N_M$, $N_U$,  the purity of the involved states and the target fidelity.
Finally, in Subsection \ref{sec:resampl}, we discuss resampling techniques which allow to determine the statistical uncertainty of the estimated fidelity in an experiment.

\subsection{Fidelities from global random unitaries}
\label{sec:fid_global}

In the main text, we present a protocol to estimate the overlap of two density matrices $\tr[]{\rho_1 \rho_2}$ using local random unitaries of the form $U_A=\bigotimes_{k\in A} U_k$ with $U_k$ from a unitary $2$-design defined on the local Hilbert space $\mathbb{C}^d$. Alternatively, one can use global random unitaries $U_A$ from a unitary $2$-design defined on the entire Hilbert space $\mathcal{H}=(\mathbb{C}^d)^{N_A}$ with dimension $D_A=d^{N_A}$. It has been proposed to prepare such global random unitaries in many-body quantum systems \cite{Nakata2017,Elben2018,Vermersch2018}.
For such global random unitaries, we obtain \cite{Elben2019}
\begin{align}
\tr[]{\rho_1 \rho_2} =D_A\sum_{\vec{s}_A,\vec{s}_A'} (-D_A)^{-D_G[\vec{s}_A,\vec{s}_A']}  \; \overline{  P^{(1)}_U(\vec{s}_A) P^{(2)}_U(\vec{s}_A') } 
\label{eq:ovg}
\end{align}
where  the ‘‘global’’ Hamming distance is  defined as  {$D_G[\vec{s}_A,\vec{s}_A']=0$} if $\vec{s}_A=\vec{s}_A'$ and  $D_G[\vec{s}_A,\vec{s}_A']=1$ if $\vec{s}_A\neq\vec{s}_A'$ and $\overline{\dots\vphantom{h}}$ denotes the ensemble average over the $2$-design on $\mathcal{H}$.
Using $1 - P^{(1)}_U(\mathbf{s}_A) = \sum_{\mathbf{s}_A\neq \mathbf{s}_A'} P^{(1)}_U(\mathbf{s}_A') $, we can identify $\Fgm$ ($\Fmax$) with the Pearson correlation coefficient \cite{Hotelling1953} (max-normalized correlation coefficient) of occupation probabilities
\begin{align}
\Fgm = &\frac{\operatorname{Cov}_U\left(P^{(1)}_U(\mathbf{s}_A) , P^{(2)}_U(\mathbf{s}_A)\right)}{\sqrt{\operatorname{Var}_U\left(P^{(1)}_U(\mathbf{s}_A)\right)\operatorname{Var}_U\left(P^{(2)}_U(\mathbf{s}_A)\right)}} \nonumber  \\&+\mathcal{O}(D_A^{-1}) \label{eq:fgm_glob}\\
\Fmax =& \frac{\operatorname{Cov}_U\left(P^{(1)}_U(\mathbf{s}_A) , P^{(2)}_U(\mathbf{s}_A)\right)}{\operatorname{max}\left[\operatorname{Var}_U\left(P^{(1)}_U(\mathbf{s}_A)\right),\operatorname{Var}_U\left(P^{(2)}_U(\mathbf{s}_A)\right) \right]} \nonumber \\&+\mathcal{O}(D_A^{-1}) . \label{eq:fmax_glob}
\end{align}
Here, we denote $\operatorname{Cov}_U\left(P^{(1)}_U(\mathbf{s}_A) , P^{(2)}_U(\mathbf{s}_A)\right) = \overline{ P^{(1)}_U(\mathbf{s}_A)P^{(2)}_U(\mathbf{s}_A)} - \overline{ P^{(1)}_U(\mathbf{s}_A)}\,\overline{P^{(2)}_U(\mathbf{s}_A)}$
for any basis state $\mathbf{s}_A$ and $\operatorname{Var}_U\left(P^{(i)}_U(\mathbf{s}_A)\right) = \operatorname{Cov}_U\left(P^{(i)}_U(\mathbf{s}_A) , P^{(i)}_U(\mathbf{s}_A)\right)$ for $i=1,2$. While in principle the above expressions appear to depend on $\mathbf{s}_A$, the averages over global random unitaries remove any dependency on the choice of basis, such that $\Fgm(\mathbf{s}_A) = \Fgm$, and $\Fmax(\mathbf{s}_A) = \Fmax$.

\paragraph*{Statistical errors for global random unitaries --}

\begin{figure}[t]
	\centering
	\includegraphics[width=0.95\linewidth]{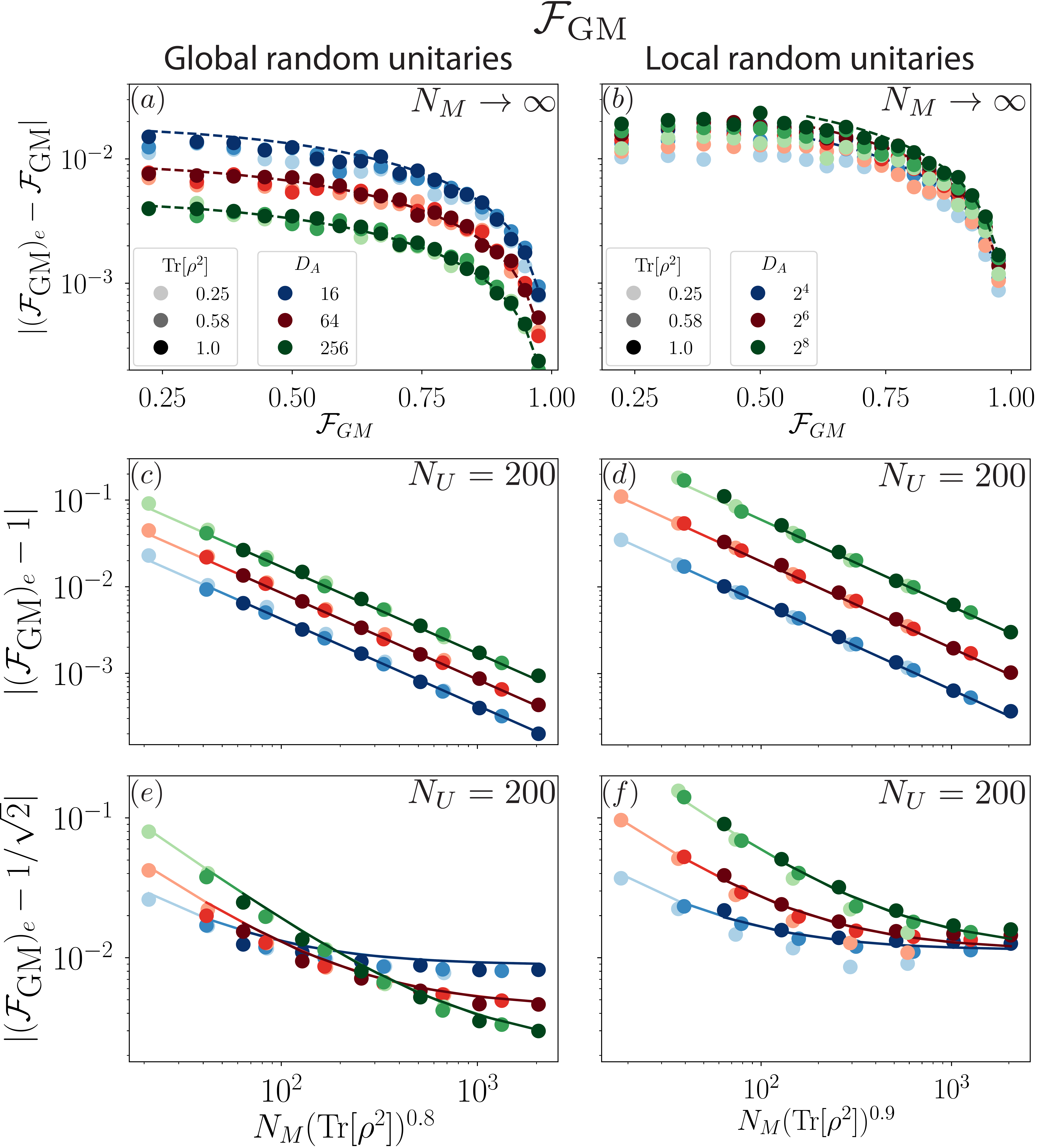}
	\caption{\textit{Statistical errors $\Fgm$.} (a,b) Scaling of the statistical error with the target fidelity itself, for $N_M\rightarrow\infty$ and $N_U=200$. (c-f) Scaling of the statistical error with the number of measurements $N_M$, (c,d) for identical product states (target fidelity unity)  and (e,f) product states with target fidelity $1/\sqrt{2}$ . In (a,c,e) global random unitaries, in (b,d,f) local random unitaries have been used.
		In all panels,  colors indicate different Hilbert space dimensions $D_A=16,64,256$, and the purity $p_2=\tr{\rho_{1,A}^2}=\tr{\rho_{2,A}^2}$ of the states increases with the darkness of the colors $p_2=0.25,0.58,1$. Mixed states $\rho_{i,A}$ ($i=1,2$) are obtained from pure product states $\ket{\psi_{i,A}}$ of $N_A=\log_2D_A$ qubits, by adding global depolarizing noise ($ \ket{\psi_{i,A}} \rightarrow \rho_{i,A}=\lambda \ketbra{\psi_{i,A}}{\psi_{i,A}} +(1-\lambda) \id/D_A$) with $\lambda=\lambda(p_2)$ chosen such that $p_2=\tr{\rho_{1,A}^2}=\tr{\rho_{2,A}^2}$. Lines are obtained from the scaling laws given in the text.    }
	\label{fig:StatErrors_GM}
\end{figure}

\begin{figure}[t]
	\centering
	\includegraphics[width=0.95\linewidth]{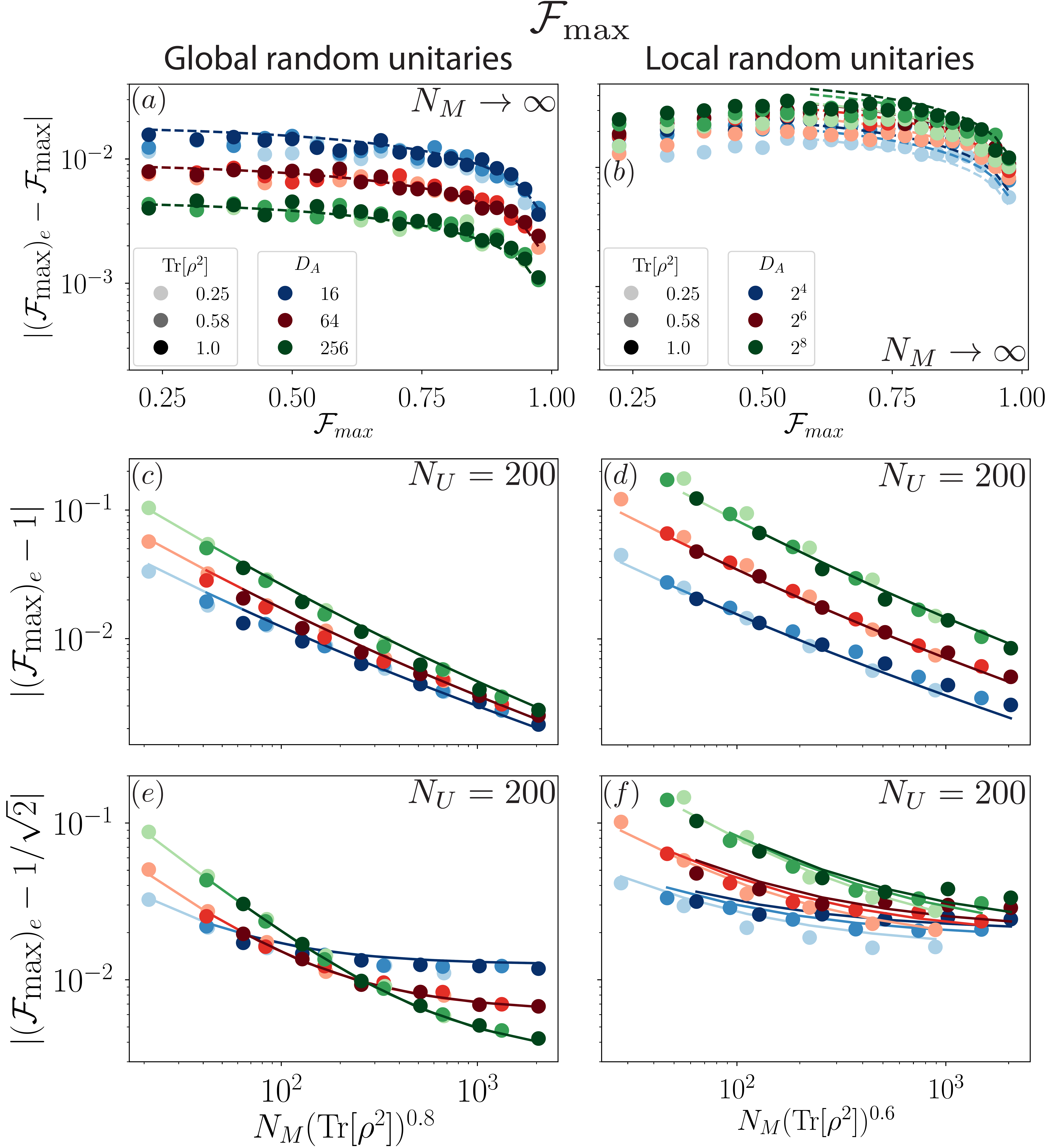}
	\caption{\textit{Statistical errors $\Fmax$}. (a,b) Scaling of the statistical error with the target fidelity itself, for $N_M\rightarrow\infty$ and $N_U=200$. (c-f) Scaling of the statistical error with the number of measurements $N_M$, (c,d) for identical product states (target fidelity unity)  and (e,f) product states with target fidelity $1/\sqrt{2}$ . In (a,c,e) global random unitaries, in (b,d,f) local random unitaries have been used.
		In all panels,  colors indicate different Hilbert space dimensions $D_A=16,64,256$, and the purity $p_2=\tr{\rho_{1,A}^2}=\tr{\rho_{2,A}^2}$ of the states increases with the darkness of the colors $p_2=0.25,0.58,1$. Mixed states $\rho_{i,A}$ ($i=1,2$) are obtained from pure product states $\ket{\psi_{i,A}}$ of $N_A=\log_2D_A$ qubits, by adding global depolarizing noise ($ \ket{\psi_{i,A}} \rightarrow \rho_{i,A}=\lambda \ketbra{\psi_{i,A}}{\psi_{i,A}} +(1-\lambda) \id/D_A$) with $\lambda=\lambda(p_2)$ chosen such that $p_2=\tr{\rho_{1,A}^2}=\tr{\rho_{2,A}^2}$. Lines are obtained from the scaling laws given in the text.  
	}
	\label{fig:StatErrors_max}
\end{figure}

We discuss first the statistical error arising  from a finite number  of random unitaries  $N_U$ used to estimate the correlation coefficients~\eqref{eq:fgm_glob} and \eqref{eq:fmax_glob}. For now, we assume  that the occupation probabilities $P^{(1)}_U(\mathbf{s}_A)$, $P^{(2)}_U(\mathbf{s}_A)$ for a given random unitary are exactly known ($N_M \rightarrow \infty$) and discuss the influence of projection noise below.
We note that,  for $D_A\gg 1$, the probabilities $P^{(i)}_U(\mathbf{s}_A)$  and $P^{(i)}_U(\mathbf{s}'_A)$ for different $\mathbf{s}_A \neq \mathbf{s}_A'$ are approximately uncorrelated.  Since we obtain in an experiment  $P^{(1)}_U(\mathbf{s}_A)$ and $P^{(2)}_U(\mathbf{s}_A)$ for all basis states $\mathbf{s}_A$ from the same experimental data, this leads to an effective sample size of $N_U D_A$ to estimate the correlation coefficients~\eqref{eq:fgm_glob} and \eqref{eq:fmax_glob}. Using 
the standard deviation of the sample distribution of the Pearson correlation coefficient \cite{Hotelling1953}, we thus find 
\begin{align}
|(\Fgm)_e - \Fgm| \sim \frac{1-\Fgm^2}{\sqrt{N_U D_A}} + \mathcal{O}\left(D_A^{-1}\right) \;. 
\label{eq:FgmNU}
\end{align}
This agrees well with numerical results, presented in Fig.~\ref{fig:StatErrors_GM}(a), which are obtained from simulating many experiments and calculating the average statistical error. 

The sampling distribution of the $\max$-normalized correlation coefficient [Eq.~\eqref{eq:fmax_glob}] is not known.  Numerically, we  find in the regime $N_U\gg1$ a similar scaling law  for the statistical error of the $\Fmax$ fidelity
\begin{align}
|(\Fmax)_e - \Fmax| \sim \frac{\sqrt{1-\Fmax^2}}{\sqrt{N_U D_A}} + \mathcal{O}\left(D_A^{-1}\right)
\label{eq:FmaxNU}
\end{align}
which is shown in Fig.~\ref{fig:StatErrors_max}(a).

In practice,  the occupation probabilities $P^{(1)}_U(\mathbf{s}_A)$ and $P^{(2)}_U(\mathbf{s}_A)$ are not known exactly, but one uses a finite number $N_M$ of measurements to estimate them. The scaling of the total statistical error (arising from a finite $N_M$ and $N_U$) with $N_M$ is shown in Figs.~\ref{fig:StatErrors_GM}(c,e) and \ref{fig:StatErrors_max}(c,e) for various target fidelities and purities of the individual product states. Overall, we find numerically that the scaling of statistical errors (for product states of purity $\tr{\rho_{A,1}^2}=\tr{\rho_{A,2}^2}=p_2$) is consistent with
\begin{align}
&|(\Fgm)_e - \Fgm| \nonumber\\&\sim \frac{1}{\sqrt{N_U D_A}} \left( 1-\Fgm^2 + c_{\text{GM}} \frac{D_A}{N_M p^{\chi_{\text{GM}}}_2} \right)\;
\end{align}
with $c_{\text{GM}}=\mathcal{O}(1)$, $\chi_{\text{GM}}\lesssim 1$ and 
\begin{align}
&|(\Fmax)_e - \Fmax|\nonumber \\& \sim \frac{1}{\sqrt{N_U D_A}} \left[ \sqrt{1-\Fmax^2} +c_{\text{max}}  \frac{D_A}{N_M p_2^{\chi_{\text{max}}}} \right. \nonumber \\
& \left.  \quad \qquad \qquad +\mathcal{O}\left(\sqrt{\frac{D_A}{N_Mp_2^{\chi_{\text{max}}}}}\right) \right]\;. 
\end{align}
with $c_{\text{max}}=\mathcal{O}(1)$, $\chi_{\text{max}}\lesssim 1$.
To summarize, we find that in order to estimate a fidelity up to an error of order $1/\sqrt{N_U}$, one needs of the order of $N_M\sim\sqrt{D_A}/p_2$ measurements.

\subsection{Local random unitaries}
\label{sec:fid_local}

For local random unitaries, the probabilities $P^{(i)}_U(\mathbf{s}_A)$ ($i=1,2$) are not independent for different $\mathbf{s}_A$ and the fidelities are  functions of  the probabilities $P^{(i)}_U(\mathbf{s}_A)$ for all basis states $\mathbf{s}_A$ [see Eq.~(2) of the MT]. We thus rely on numerical simulations of many experiments to obtain the average statistical error.
We find in the regime $\Fmax=\mathcal{O}(1)= \Fgm$ that for product states scaling laws of the form
\begin{align}
& |(\Fgm)_e - \Fgm|\nonumber \\ & \sim \frac{1}{\sqrt{N_U }} \left( c^{(1)}_{\text{GM}}(1-\Fgm^2) + c^{(2)}_{\text{GM}} \frac{D_A^{0.8}}{N_M p^{\chi_{\text{GM}}}_2} \right)\;
\end{align}
with  $c^{(1)}_{\text{GM}},c^{(2)}_{\text{GM}}=\mathcal{O}(1)$ and $\chi_{\text{GM}}\lesssim 1$ and
\begin{align}
&|(\Fmax)_e - \Fmax|\nonumber \\& \sim \frac{1}{\sqrt{N_U}} \left[ c^{(1)}_{\text{max}} \sqrt{1-\Fmax^2} + c^{(2)}_{\text{max}} \frac{D_A^{0.8}}{N_M p^{\chi_{\text{max}}}_2}\right. \nonumber \\
&\qquad  \qquad \left.+\mathcal{O}\left( \sqrt{\frac{D_A^{0.8}}{N_Mp^{\chi_{\text{max}}}_2}}\right) \right]
\end{align}
with    $c^{(1)}_{\text{max}}, c^{(2)}_{\text{max}}=\mathcal{O}(1)$ and $\chi_{\text{max}}\lesssim 1$ are consistent with numerical results [see Figs.~\ref{fig:StatErrors_GM} and \ref{fig:StatErrors_max} panels (b,d,f)].
We thus find  that the statistical error with local random unitaries is, compared to global random unitaries and in the limit $N_M\to \infty$ by a factor $1/\sqrt{D_A}$ larger. This is expected since for local random unitaries the $P^{(i)}_U(\mathbf{s}_A)$ for various $\mathbf{s}_A$ are not independent and the effective sample size to estimate the fidelities is thus just given by $N_U$. 
In summary, in a typical experimental situation, projection noise caused by finite $N_M$ is the dominating source of statistical errors. Irrespective of the target fidelity, the required number of projective measurements is of the order $N_M\sim 2^{bN_A}/p_2$ with $b\lesssim 1$ (see also MT).

\subsection{Resampling techniques and allocation  of the measurement budget}
\label{sec:resampl}
\begin{figure}[t]
	\centering
	\includegraphics[width=0.95\linewidth]{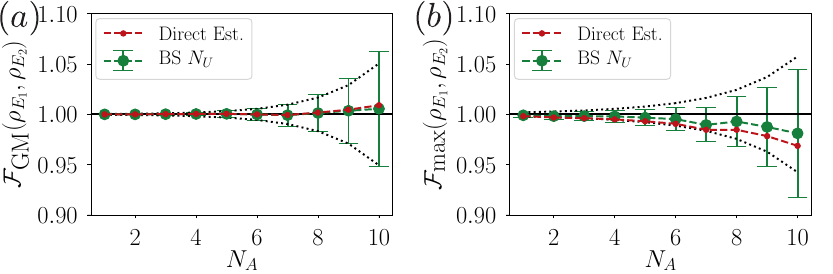}
	\caption{\textit{Bootstrap Resampling.} (a,b) Fidelities $\Fmax$ and $\Fgm$ for pure product states with unit target fidelity as a function of the number of qubits $N_A$  for $N_U=250$ and $N_M=400$. Red lines are direct (biased) estimations of the fidelities,  dotted lines are bias corrected estimations obtained with Bootstrap resampling across the random unitaries, see text for details. Error bars are statistical errors,  obtained from Bootstrap resampling across unitaries. Each quantity is a mean over 100 simulated experiments. Black lines  indicate the expected statistical errors calculated with the scaling laws  Eqs.~(C8) and (C9), drawn above and below unit target fidelity.  }
	\label{fig:resampling}
\end{figure}

In an experiment, one would like to infer the statistical uncertainty of the measured fidelities directly from the measured data (and not from performing the experiment many times with the same parameters $N_M$ and $N_U$). To achieve this, we use Bootstrap resampling \cite{Efron1983} across the random unitaries to estimate the standard error. Typically, we draw, with replacement, $250-500$ Bootstrap re-samples, each of size $N_U$. We take the data of the  projective measurements corresponding to the unitaries in each re-sample ($N_M$ projective measurement per unitary) and estimate the fidelity for each re-sample. Then, the Bootstrap estimate of the standard errors is just given by the standard error of the mean of the set of newly estimated fidelities \cite{Efron1983}. In Fig.~\ref{fig:resampling}, we show that the standard errors (error bars) obtained with such Bootstrap resampling are of a similar size as the average statistical errors obtained from simulating many experiments (indicated through black lines). We thus conclude that Bootstrap resampling allows us to infer the statistical uncertainty.

To infer the fidelities, we estimate the overlap and purity and subsequently calculate the fidelities using Eq.~(2) of the MT. This procedure results in general, for finite $N_M$ and $N_U$, in a biased estimation of the fidelities, with a bias scaling with the statistical errors of purity and overlap.  In Fig.~\ref{fig:resampling}, we show numerically that  the estimation of $\Fmax$ ($\Fgm$) is indeed biased towards lower (higher) values (red line), with a bias which is, for experimentally relevant parameters, of the order of a few percent. Bootstrap resampling allows  corrections for such a bias to first order in $1/{N_U}$ (Fig.~\ref{fig:resampling} green dotted lines) \cite{Efron1983}. Here, the Bootstrap estimate of the bias is given as the difference of the mean estimated fidelity over the bootstrap resamples and the original estimation.  Using these Bootstrap estimates, we present first order unbiased estimators in all plots showing experimental data.

Finally, we note that the standard error estimation with bootstrapping is the basic ingredient for an  algorithm to choose iteratively, based on the already obtained data, the allocation of the total measurement budget $N_UN_M$ into random unitaries $N_U$ and projective measurements per unitary $N_M$ in an experiment. The procedure is as follows: Initially, one performs experimentally  the fidelity estimation  with a small number of unitaries and measurements per unitary, $N_U\approx N_M\approx50$, and uses bootstrapping to infer its standard error. Then, the experimental data of either $n\approx 10$ unitaries ($N_U\rightarrow N_U-n$)  or $n\approx 10$ measurements per unitary ($N_M\rightarrow N_M-n$) is removed, and  the fidelity is estimated two more times,  from both reduced data sets. The standard error of the two new estimations, obtained from bootstrapping on the reduced data set,  is compared to the standard error of original estimation. From the direction ($N_U\rightarrow N_U-n$ or $N_M\rightarrow N_M-n$) where the standard error increases most, one can expect the strongest reduction if $N_U$ or $N_M$ is increased, respectively. Consequently, more experimental runs are performed, either with the same unitaries as before but with increased $N_M\rightarrow N_M +n$ or with more unitaries $N_U\rightarrow N_U +n$ but constant number of measurements per unitary. This procedure is repeated iteratively until the standard error of the estimated fidelity has decreased to a predefined value.  

\section{Experiment-Theory fidelity estimation}
\label{app:ExperimentTheory}

By replacing one of the two quantum devices with a classical computer simulating the experiment, we can use the presented protocol to measure an experiment-theory fidelity towards a known theoretical target state. On the theory side, one applies the same random unitaries $U_A$ as in the experiment to the classical representation of the target state $\rho_T$, and determines the outcome probabilities $P_U^{T}( \mathbf{s}_A)$. Together with the $P_U^{E}( \mathbf{s}_A)$ estimated from the experimental data, one obtains the experiment-theory fidelity $\Fmax(\rho_E,\rho_T)$ [or  $\Fgm(\rho_E,\rho_T)$] via Eq.~(2) of the MT. With this procedure we obtain, using the data from Ref.~\cite{Brydges2019}, experiment-theory fidelities for (sub)systems of up to $10$ qubits (see Figs.~3 (MT), \ref{fig:ExpExpGM} and \ref{fig:unconcon}). 

\begin{figure}[t]
	\centering
	\includegraphics[width=0.95\linewidth]{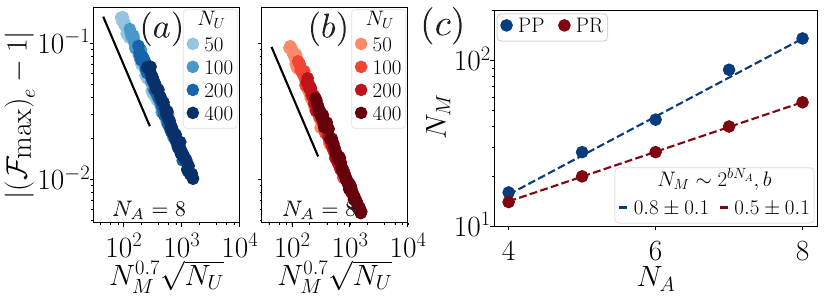}
	\caption{\textit{Scaling of the required number of measurements for experiment-theory estimation.} (a,b) Average statistical error $|(\Fmax(\rho_A,\rho_A))_e -1 |$  as a function of the number of measurements $N_M$ per random unitary for various $N_U$ (darkness of colors). The state $\rho_A$ of $N_A=8$ qubits ($d=2$)  is taken to be  (a) a pure product state (PP)  and (b) a pure Haar random state (PR).  Black lines are guides for eye $\sim 1/(N_M^{0.7} \sqrt{N_U})$. (c) Scaling of the minimal number of required measurements $N_M$  to estimate $(\Fmax(\rho_E,\rho_T))_e$ up to a fixed   statistical error of $0.05$ as a function of the number of qubits $N_A$, for fixed $N_U=100$. }
	\label{fig:StatErrorExTheory}
\end{figure}

In the following, we analyze the statistical errors of the experiment-theory fidelity estimation. Compared to the experiment-experiment scenario,   the probabilities  $P_U^{T}( \mathbf{s}_A)$  on the theory side are typically exactly known, and thus shot-noise affects only $P_U^{E}( \mathbf{s}_A)$. In Fig.~\ref{fig:StatErrorExTheory}(a,b), we thus observe a smaller overall statistical error (compared to the results in the MT) which scales as $~1/(N_M^{0.7}\sqrt{N_U})$.
In Fig.~\ref{fig:StatErrorExTheory}(c), we display the scaling of the   number of required measurements $N_M$ ($N_U=50$) to obtain the fidelity up to an error $\epsilon=0.05$ as a function of the number of qubits. We observe similar scaling exponents as for the experiment-experiment fidelity estimation. We attribute this to the fact that shot-noise arising from the finite number of measurements $N_M$ performed in the experiment per random unitary is the dominating source of error. 

We remark that the experiment-theory estimation, for pure theory states, can also be achieved using direct fidelity estimation \cite{DaSilva2011,Flammia2011}. Here, the knowledge of the theory state is explicitly taken into account to achieve an efficient fidelity estimation for well-conditioned states with support on a few multi-qubit Pauli operators. We expect that taking into account (partial) knowledge about the quantum states of interest can also decrease the required number of measurements for the presented method which is subject to future work.

\section{Experimental imperfections}
\label{app:Imperfections}

In this section, we calculate and experimentally investigate the effect of systematic errors in our protocol arising from a mismatch of the random unitaries applied in both devices and from decoherence during the application of the random unitaries.
\subsection{Modelling of errors}
\label{sec:ModelErrors}
Our approach to estimate cross-platforms fidelities is based on realizing the \emph{same} local random unitaries $U_A$ on two different platforms. 
Restricting for clarity to the case of qubits ($d=2$), we study the effects of systematic errors due to:

(i) Unitary errors in the realization of random unitaries, e.g., due to small miscalibration of the quantum hardware. In order to model this effect, we assume that, instead of a random unitary $U_A$,  each device implements a  random unitary of the form $U_A^{(i)}=U_AV_A^{(i)}$ with $V_A^{(i)}=\bigotimes_{k=1}^{N_A }\exp(i h^{(i)}_k \eta_i)$ 
with $h^{(i)}_k$ being a random Hermitian matrix sampled for all $i,k$ independently from the \textcolor{black}{Gaussian unitary ensemble \footnote{This is an ensemble of Hermitian random matrices with complex entries which  real and imaginary parts are independently distributed according to the standard normal distribution \cite{Haake2010}.}}. Here, $\eta_i$ quantifies the level of imperfection, with $\eta_i=0$ corresponding to perfect operations.
In particular, we assume thus that the erroneous additional rotation $V^{(i)}$ is a local random unitary, which is independent of $U$.

(ii) In addition, we consider the presence of decoherence acting during the application of the local random unitaries, which is modeled as local depolarization of the the form of:
\begin{eqnarray}
\tilde \rho_{i,A} &=& (1-{\frac{3}{2}p_{D,i} N_A}) \rho_{i,A} + \frac{p_{D,i}}{2} \sum_{k,\gamma}\sigma_k^\gamma \rho_{i,A} \sigma_k^\gamma \label{eq:deph}
\\
&=& (1-{2p_{D,i}N_A}) \rho_{i,A} +2 {p_{D,i}} \sum_{k} \mathrm{Tr}_k (\rho_{i,A})\otimes \frac{ \mathbf{1}_k}{2},
\nonumber 
\end{eqnarray}
with $k=1,\dots,N$, $\gamma=x,y,z$, and $p_{D,i}\ll 1$ the single qubit decoherence error  for platform $i$. \textcolor{black}{Equation \eqref{eq:deph}} supposes that each qubit on platform $i$ can be projected to the identity matrix, i.e., `depolarized', with probability $2p_{D,i}$.

In the presence of (i) unitary errors and (ii) depolarization, we infer $\mathrm{Tr}(\rho_1 \rho_2)$ and purities thus in an experiment from estimators of the form
\begin{align}
E_{i,j}\!=\! \label{eq:imperfectE} 
2^{N_{A}}\!\sum_{\mathbf{s}_{A},\mathbf{s}_{A}'}\!(-2)^{-\mathcal{D}[\mathbf{s}_{A},\mathbf{s}_{A}']}\;\overline{\tilde  P_{U^{(i)}}^{(i)}(\mathbf{s}_{A})\tilde  P_{U^{(j)}}^{(j)}(\mathbf{s}_{A}')}
\end{align}
with $ \tilde P_{U^{(i)}}^{(i)}(\mathbf{s}_{A}) = \tr[]{U^{(i)}_A \tilde \rho_{i,A}( U^{(i)}_A )^\dagger \ketbra{\mathbf{s}_A}{\mathbf{s}_A}}$.

\subsection{Error estimates}
\label{sec:EstimateErrors}

In the following, we evaluate the estimators $E_{i,j}$. For simplicity of notation, we drop the subscript $A$ in this subsection.
Due to the independence of  $V^{(i)}$ and $U$, we can use Eq.~(2) MT, and find 
\begin{align}
E_{1,2} &= \tr[]{\overline{V^{(1)} \tilde \rho_{1} V^{(1)\dag}}  \, \overline{V^{(2)} \tilde \rho_{2} V^{(2)\dag}}} \label{eq:Estimator12}
\end{align}
and
\begin{align}
E_{i,i} &= \tr[]{ \tilde{\rho}_i   \tilde{\rho}_i } \quad \text{for} \quad i=1,2 .
\label{eq:Estimator11}
\end{align}
A mismatch between the random unitaries applied in both experiments due to unitary errors (i) thus affects the estimation $E_{1,2}$ of the overlap, but does not affect the purity estimation $E_{i,i}$.

We now evaluate $E_{1,2}$ [Eq.~\eqref{eq:Estimator12}] using  the assumption that $V^{(i)}=\bigotimes_{k=1}^{N_A }\exp(i h^{(i)}_k \eta_i)$ 
with $h^{(i)}_k$ being a random Hermitian matrix sampled for all $i,k$ independently \textcolor{black}{ from the Gaussian unitary ensemble \cite{Haake2010}}. In particular, we use the property
\begin{align}
\overline{[h^{(i)}_k]_{a,b} [h^{(i)}_k]_{c,d} } &= \delta_{a,d}\delta_{b,c},\label{eq:1design}
\end{align}
with $\delta$ the Kronecker delta and $\overline{\dots \vphantom{h}}$ the ensemble average.
We expand Eq.~\eqref{eq:Estimator12} to leading order in $\eta_i$, use  the independence between random matrices $h^{(i)}_k$ and Eq.~\eqref{eq:1design} and find
\begin{align}
E_{1,2} =& \tr[]{\tilde \rho_1  \tilde \rho_2}
\nonumber \\
&+ \sum_{\substack{i,j=1,2\\i \neq j}} \eta_i^2  \sum_k \tr[]{\tilde\rho_{i} \overline{h^{(j)}_k  \tilde\rho_j h^{(j)}_k  }}
\nonumber \\
&- \sum_{\substack{i,j=1,2\\i \neq j}} \frac{\eta_i^2}{2} \sum_k  \tr[]{\tilde\rho_{i}  \overline{h^{(j),2}_k}   \tilde\rho_j+ \tilde\rho_{i}  \tilde\rho_j \overline{h^{(j),2}_k}  } 
\nonumber \\ 
=&[1-2 (\eta_1^2 + \eta_2^2) N ] \; \tr[]{\tilde\rho_1  \tilde\rho_2}
\nonumber \\ 
&+ (\eta_1^2+\eta_2^2)  \sum_k  \tr[]{\tr[\{k\}]{\tilde\rho_1} \tr[\{k\}]{ \tilde\rho_2}},  \label{eq:E12}
\end{align}
where $\tr[\{k\}]{\tilde\rho_1}$ denotes the partial trace over qubit $k$.

\begin{figure}[t]
	\centering
	\includegraphics[width=0.99\linewidth]{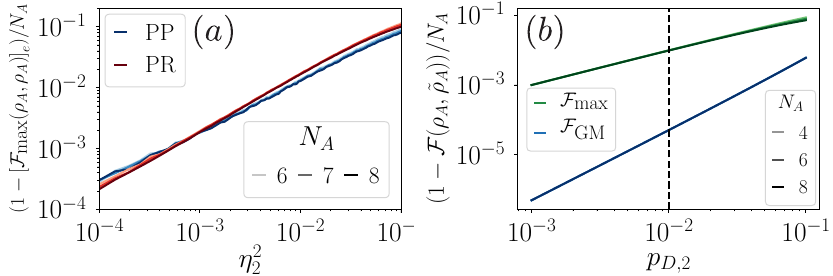}
	\caption{\textit{Robustness against imperfections.} (a) Influence of unitary errors in the implementation of random unitaries, simulated on a classical computer. Shown is the error of the estimated fidelity as function of the error strength $\eta_2^2$ for two types of states $\rho_{1,A}=\rho_{2,A}=\rho_A$ of $N_A$ qubits: PP is a pure product state and PR a pure Haar random state. We use $N_U=500$ unitaries to extract the fidelity. These unitaries are imperfect on platform 2, i.e.\ modified with random over/underrotation with strength $\eta_2$ (see text). Projection noise is not included ($N_M\rightarrow \infty$).  (b) Error of the estimated fidelity as a function of the strength $p_{D,2}$ of local depolarizing noise, acting during the application of  random unitaries in platform $2$, $\rho_{A,2}\rightarrow \tilde{\rho}_{A,2}$(see text). While here   PP states  with unit target fidelity $\rho_{A,1}=\rho_{A,2}$ are shown, PR states behave identically. Black dashed line indicates a typical experimental value, taken from \cite{Brydges2019}.}
	\label{fig:Errors}
\end{figure}

We now analyze the effect of decoherence (ii). To this end, we evaluate
\begin{align}
\tr[]{\tilde\rho_i \tilde\rho_{j}}=& \left(1-{2\left[p_{D,i}+p_{D,j}\right]N} \right) \tr[]{\rho_i \rho_{j} } 
\\ 
&+{(p_{D,i}+p_{D,j})}  \sum_k  \tr[]{ \tr[\{k\}]{ \rho_i}  \tr[\{k\}]{ \rho_j} }.\nonumber  \label{eq:trrhot}
\end{align}
Thus, decoherence affects both the estimation of the overlap $E_{1,2}$ and of the purities $E_{i,i}=\tr[]{\tilde\rho_i \tilde\rho_{i}}$.
We can now calculate the estimators of the two fidelities
\begin{eqnarray}
\tilde{\mathcal{F}}_{GM}&=&\frac{E_{1,2}}{\sqrt{E_{1,1}E_{2,2}}} 
\nonumber \\
\tilde{\mathcal{F}}_{\max}&=&\frac{E_{1,2}}{\max(E_{1,1},E_{2,2})}. 
\end{eqnarray}
and extract the relevant errors for certain quantum states $\rho_1$ and $\rho_2$. Let us consider here for illustration the situation $\rho_1=\rho_2$ corresponding to unit fidelities. In this case, the expressions simplify  considerably:
\begin{align}
\tilde{\mathcal{F}}_{GM}=&\mathcal{F}_{GM} \nonumber \\
&- 2 (\eta_1^2 + \eta_2^2) N + (\eta_1^2+\eta_2^2)  \sum_k  \frac{ \tr{\tr[\{k\}]{\rho_1}^2}}{\tr[]{\rho^2_1}}\nonumber \\
&+ \mathcal{O}(p_{D,1}^2, p_{D,2}^2, \eta_1^4,\eta_2^4, p_{D,2}\eta_2^2, p_{D,1}\eta_1^2).
\end{align}
For $\Fmax$ we consider the case $E_{1,1}> E_{2,2}$, then
\begin{align}
\tilde{\mathcal{F}}_{\max}=&\mathcal{F}_{\max} \nonumber  - {2p_{D,2}N}+ {p_{D,2}}  \sum_k  \frac{ \tr{\tr[\{k\}]{\rho_1}^2}}{\tr[]{\rho^2_1}}
\nonumber \\
&- 2 (\eta_1^2 + \eta_2^2) N 
+ (\eta_1^2+\eta_2^2)  \sum_k  \frac{ \tr{\tr[\{k\}]{\rho_1}^2}}{\tr[]{\rho^2_1}} \nonumber\\
&+ \mathcal{O}(p_{D,1}^2, p_{D,2}^2, \eta_1^4,\eta_2^4, p_{D,1}\eta_1^2, p_{D,2}\eta_2^2). 
\end{align}

In summary, both fidelities are affected with  $\eta^2_i$ in miscalibration errors, while only the  fidelity $\Fmax$ can suffer from depolarization in first order in $p_{D,i}$. This is in agreement with the numerical study presented in Fig.~\ref{fig:Errors}. 
Here, we take the two states $\rho_1=\rho_2$ to be identical and estimate the fidelity using the outlined protocol, but with imperfect random unitaries. In panel (a), we present the average error of the estimated fidelity $\Fmax$ as a function of the strength $\eta^2_2$ of unitary errors of the form (i) present in platform 2 ($\eta_1=0$). In agreement with the analytical results, we find a linear increase of the error of the estimated fidelity with increasing $\eta^2_2$, for both, $\rho_1=\rho_2$ beeing pure product states (PP) and pure random states (PR). In panel (b), we show the error of the estimated fidelities $\Fgm$ and $\Fmax$ as a function of the strength $p_{D,2}$ of the local depolarization noise present in platform 2,  $p_{D,1}=0$.  As expected, we find a linear (quadratic) increase of the error of the estimated $\Fmax$ ($\Fgm$) with increasing $p_{D,2}$. In both panels (a,b), we further find the error to be proportional to $N_A$. Finally, we emphasize that both, unitary errors and local depolarization, lower the estimated fidelity and do not lead to false positives.

\subsection{Testing experimentally the influence of imperfections}

In this subsection, we analyze the influence of imperfections in our protocol experimentally.
\begin{figure}[H]
	\centering
	\includegraphics[width=0.99\linewidth]{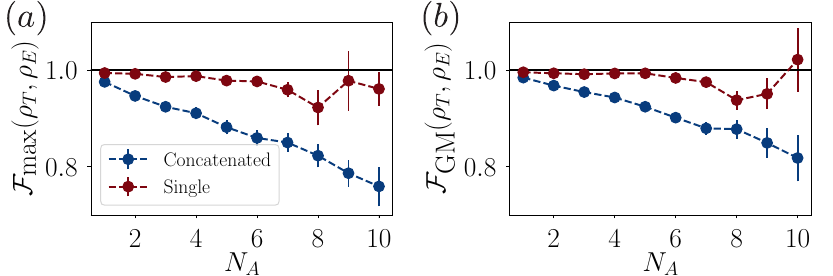}
	\caption{\textit{Testing experimentally the influence of imperfections.} Estimated Experiment-Theory fidelities \Fmax (a) and \Fgm (b) as functions of the subsystem size $[1\rightarrow N_A]$ for fully polarized product states, i.e.\ $\rho_T=\ket{\Psi_0}\bra{\Psi_0}$ with $\ket{\Psi_0}=\ket{00\dots 0}$. Total system size is $N=10$ qubits. Each randomized measurement has been experimentally implemented with a single random unitary (red) or two concatenated random unitaries (blue).
		In both panels, $N_U=500$ and $N_M=150$.}
	\label{fig:unconcon}
\end{figure}
We prepare a fully polarized product state $\ket{\Psi_0}=\ket{00\dots 0}$, and aim to estimate experimentally its fidelity to the known theoretical target. To characterize imperfections, we implement each randomized measurements first with a single random unitary per qubit and secondly with  concatenating two random unitaries per qubit.

We find that  the fidelities which are estimated with two concatenated unitaries are significantly smaller than the ones estimated with  a single random unitary [Fig.~\ref{fig:unconcon}]. The difference for $\Fgm$ is less pronounced than for $\Fmax$. This is consistent with the error model presented in Secs.~\ref{sec:ModelErrors}  and~\ref{sec:EstimateErrors}: Both unitary errors and decoherence increase when concatenating two random unitaries, and the estimated fidelities thus decrease. The fidelity $\Fgm$ is robust against decoherence to first order, and thus less affected.
We emphasize that the systematic errors  discussed in this section, unitary errors in the preparation and decoherence during the application of the random unitaries, decrease the estimated fidelities, and thus do not lead to false positives.
We further remark that the data of Ref.~\cite{Brydges2019}  was taken using two concatenated random unitaries. Here, we show now, using the presented protocol to perform theory-experiment fidelity estimation, a clear improvement if single unitaries are used. This demonstrates the utility of the protocol to check and benchmark concrete experiments.

\end{document}